\documentclass[namedreferences]{solarphysics}
\usepackage[hyperref,natbib,solaenum,linksfromyear,optionalrh,solaromanenum]{spr-sola-addons} % For Solar Physics 
\usepackage{graphicx}                    % For eps figures, newer & more powerfull
\usepackage{color}                       % For color text: \color command
\usepackage{breakurl}                         % For breaking URLs easily trough lines
\ifx \doiurl    \undefined \def \doiurl#1{\href{http://dx.doi.org/#1}{\textsf{#1}}}\fi
\ifx \adsurl    \undefined \def \adsurl#1{\href{http://adsabs.harvard.edu/abs/#1}{\textsf{#1}}}\fi
\ifx \arxivurl  \undefined \def \arxivurl#1{\href{http://arxiv.org/abs/#1}{\textsf{#1}}}\fi
\ifx \urlurl    \undefined \def \urlurl#1{\href{http://#1}{\textsf{#1}}}\fi
\begin{document}

\begin{article}

\begin{opening}

\title{Formation of a Flare-Productive Active Region:
  Observation and Numerical Simulation of NOAA AR 11158}

%%%%%%%%%%%%%%%%%%%%%%%%%%%%%%%%%%%%%%%%%%%%%%%%%%%
%% Authors Names
%
\author{S.~\surname{Toriumi}$^{1,2}$\sep
        Y.~\surname{Iida}$^{3}$\sep
        K.~\surname{Kusano}$^{4,5}$\sep
        Y.~\surname{Bamba}$^{4}$\sep
        S.~\surname{Imada}$^{4}$
      }

%%%%%%%%%%%%%%%%%%%%%%%%%%%%%%%%%%%%%%%%%%%%%%%%%%%
%% Runningheads
%
\runningauthor{S.~Toriumi {\it et al}.}
\runningtitle{Formation of NOAA AR 11158}

%%%%%%%%%%%%%%%%%%%%%%%%%%%%%%%%%%%%%%%%%%%%%%%%%%%
%% Affilations 
%
  \institute{$^{1}$ Department of Earth and Planetary Science,
     University of Tokyo,
     Hongo, Bunkyo-ku, Tokyo 113-0033, Japan\\
     email: \url{toriumi@eps.s.u-tokyo.ac.jp}\\
     $^{2}$ JSPS Research Fellow\\
     $^{3}$ Institute of Space and Astronautical Science,
     Japan Aerospace Exploration Agency,
     Chuo-ku, Sagamihara, Kanagawa 252-5210, Japan\\
     $^{4}$ Solar-Terrestrial Environment Laboratory,
     Nagoya University, Furo-cho, Chikusa-ku, Nagoya,
     Aichi 464-8601, Japan\\
     $^{5}$ Japan Agency for Marine-Earth Science and Technology (JAMSTEC),
     Kanazawa-ku, Yokohama, Kanagawa 236-0001, Japan
   }

%%%%%%%%%%%%%%%%%%%%%%%%%%%%%%%%%%%%%%%%%%%%%%%%%%%
%%% Abstract 
\begin{abstract}
We present a comparison
of the {\it Solar Dynamics Observatory} (SDO) analysis
of NOAA Active Region (AR) 11158
and numerical simulations
of flux-tube emergence,
aiming to investigate the formation process
of the flare-productive AR.
First,
we use SDO/{\it Helioseismic and Magnetic Imager} (HMI) magnetograms
to investigate the photospheric evolution
and {\it Atmospheric Imaging Assembly} (AIA) data to analyze 
the relevant coronal structures.
Key features
of this quadrupolar region are
a long sheared polarity inversion line (PIL)
in the central $\delta$-sunspots
and a coronal arcade above the PIL.
We find that
these features
are responsible
for the production of intense flares
including an X2.2-class event.
Based on the observations,
we then propose
two possible models
for the creation of AR 11158
and conduct flux emergence simulations
of the two cases
to reproduce this AR.
Case 1
is the emergence of a single flux tube,
which is split into two
in the convection zone
and emerges at two locations,
while Case 2
is the emergence
of two isolated but neighboring tubes.
We find that,
in Case 1,
a sheared PIL and a coronal arcade
are created
in the middle of the region,
which agrees with the AR 11158 observation.
However, Case 2 never build
a clear PIL,
which deviates from the observation.
Therefore,
we conclude that
the flare-productive AR 11158 is,
between the two cases,
more likely to be created
from a single split emerging flux
than two independent flux bundles.
\end{abstract}

%%%%%%%%%%%%%%%%%%%%%%%%%%%%%%%%%%%%%%%%%%%%%%%%%%%
%% Keywords
%
\keywords{Active Regions, Magnetic Fields; Flares, Relation to Magnetic Field; Interior, Convective Zone}

\end{opening}
%-------------------------------------------------

%%%%%%%%%%%%%%%%%%%%%%%%%%%%%%%%%%%%%%%%%%%%%%%%%%%
%% Sections
%% Figure 
%% Table

\section{Introduction
  \label{sec:intro}}

Solar flares are
catastrophic eruptions
produced around
active regions (ARs).
Also, AR is the product
of emerging magnetic flux
from the deep convection zone
({\it e.g.} \opencite{fan09}).
Observations have revealed that
intense flares often occur
near polarity inversion lines (PILs)
of ARs,
especially those with strong magnetic shear
({\it e.g.} \opencite{hag84}).
This is due to
the availability
of free magnetic energy
in the sheared coronal arcades
over such PILs.
When a flare occurs,
the stored energy
is released
via magnetic reconnection
(see \opencite{shi11},
and references therein).
\inlinecite{kus12} carried out
a systematic study
of three-dimensional magnetohydrodynamic (3D MHD) simulations
to investigate the triggering mechanism
of solar flares.
It was found that
a solar flare occurs
at the highly sheared PIL
and the overlying coronal arcade
above it.
The flare is triggered
by a small-scale magnetic field
that initiates the reconnection
between the coronal arcades.
The magnitude of flare eruptions
({\it e.g.} maximum kinetic energy)
was found to increase
with a shear angle of the arcade,
which is measured
from the axis normal to the PIL.

As demonstrated by
\inlinecite{kus12},
the creation of a sheared PIL
is critical for
the production of flares.
Here,
the PIL in an AR
is formed through
the large-scale flux emergence
of the entire AR.
If the emerging flux
transports larger magnetic helicity,
it may build up a more sheared PIL
and may produce stronger flares.
If the flux is severely deformed
before it appears
at the surface,
it may develop a more complex AR
($\delta$-sunspots,
possibly with a sharp, sheared PIL),
which is known to produce
larger flares
\citep{sam00}.
Therefore,
intense flares
at a highly sheared PIL
are likely to reflect
the evolutionary history
of emerging magnetic flux
while in the convection zone
(see also \opencite{poi13}).

In this article,
we present a detailed analysis
of NOAA AR 11158
and a comparison
with numerical simulations
of emerging magnetic flux.
The aim of this
work is to study
the subsurface/global structure
of a flare-productive AR.
For this purpose,
we first analyze
observational data of AR 11158
obtained by the {\it Helioseismic and Magnetic Imager}
(HMI: \opencite{sche12}; \opencite{scho12})
and {\it Atmospheric Imaging Assembly}
(AIA: \opencite{lem12})
onboard the {\it Solar Dynamics Observatory}
(SDO: \opencite{pes12})
to investigate
the photospheric and coronal evolution
of this AR.
Thanks to the continuous,
full-disk observation by SDO,
we are able to investigate
the evolution
of this AR
from its earliest stage.
Then,
we conduct numerical simulations
of emerging flux tubes
to reproduce AR 11158.
By comparing the observational
and the numerical results,
particularly the geometrical evolution
of surface magnetic fields
around the PIL
and of overlying coronal arcades,
we search for a possible scenario
of the large-scale emerging flux
that creates a sheared PIL
in AR,
which is largely responsible
for producing strong flares.

The rest of the article
is organized
as follows:
We first describe
the observations and the simulations
in Sections \ref{sec:observation}
and \ref{sec:simulation},
respectively.
Then,
a comparison of observations and simulations
is given in Section \ref{sec:comparison}.
We discuss the results
in Section \ref{sec:discussion}
and, finally,
we summarize the article
in Section \ref{sec:conclusion}.

\section{Observation and Data Analysis
  \label{sec:observation}}

\subsection{Observations}

In this section,
we analyze the observations
of NOAA AR 11158
by the SDO spacecraft.
This flare-productive AR,
which produced the first X-class flare
of Solar Cycle 24
(\opencite{schr11}),
emerged in the southern hemisphere
in February 2011.
The entire growth
from its birth
to production of flares
was on the near side of the Sun.

We use
tracked line-of-sight magnetograms
of AR 11158,
which were obtained by SDO/HMI.
We applied Postel's projection
to the data cube;
namely,
the magnetograms were projected
as if they seen directly from above.
The data cube
has a pixel size of 0.5 arcsec
($\approx 360\ {\rm km}$)
with $512^{2}$ pixel field of view,
and a temporal resolution of
12 minutes
spanning a duration of seven days,
starting from 00:00 UT,
9 February 2011.
We also use SDO/AIA 171 \AA\, data
to investigate the coronal evolution
of this AR.
The data that we analyzed
has a pixel size of 0.6 arcsec
($\approx 435\ {\rm km}$),
and we cut out the data
from a series of full-disk images
without applying any geometrical projection.
In addition,
we use the soft X-ray flux
obtained by the 
{\it Geostationary Operational Environmental Satellite} (GOES)
to monitor solar activity.

\subsection{Evolution of AR 11158}

\subsubsection{Photospheric evolution}

\begin{figure}
  \centerline{\includegraphics[width=0.9\textwidth,clip=]{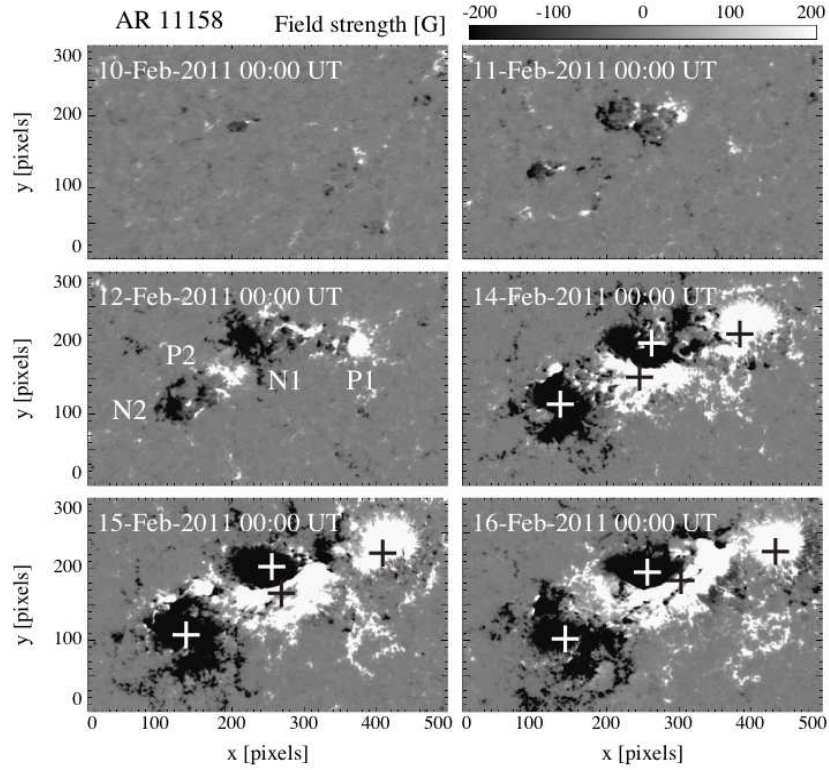}}
  \caption{SDO/HMI magnetogram of NOAA AR 11158.
    The flux-weighted centers
    of the four major polarities,
    P1, N1, P2, and N2,
    are indicated with cross signs.
    One pixel corresponds to 0.5 arcsec.
  }
  \label{fig:hmi}
\end{figure}

Figure \ref{fig:hmi} presents
the magnetic evolution of AR 11158.
This AR is basically
composed of two emerging bipoles,
which we identify
as P1--N1 and P2--N2.
The first bipole [P1--N1] appeared
in the southern hemisphere
on 9 February,
while the second pair [P2--N2]
appeared on 10 February.
In both bipoles,
the preceding (western) polarity
is positive,
which agrees with
the expected alignment
of Hale's polarity law
\citep{hal19}
for Solar Cycle 24.
After this,
each magnetic element
of both bipolar systems
gradually separated
from its counterpart.
One of the key features of this AR is
the relative motion of N1 and P2.
As time progressed,
the southern positive polarity P2
came closer to
the northern negative polarity N1
and the magnetic gradient normal to
the PIL
between N1 and P2 became steeper.
Then, from 13 February,
P2 continuously drifted to the West
and rotated around the southern edge of N1,
making the PIL highly sheared and elongated.
N1 and P2 shared
a common penumbra,
forming $\delta$-sunspots.
The length of the PIL
was about $30\ {\rm Mm}$
on 15 February.
The counterclockwise motion
of the photospheric patches N1 and P2
lasted at least until 00:00 UT, 16 February,
{\it i.e.}
the end of the observational data set
that we analyzed.
In the final stage,
P2 became elongated and moved
to the outmost positive polarity P1
as if P2 merged into P1.
At this time,
the length
of the AR
exceeded
$150\ {\rm Mm}$.

\begin{figure}
  \centerline{\includegraphics[width=1\textwidth,clip=]{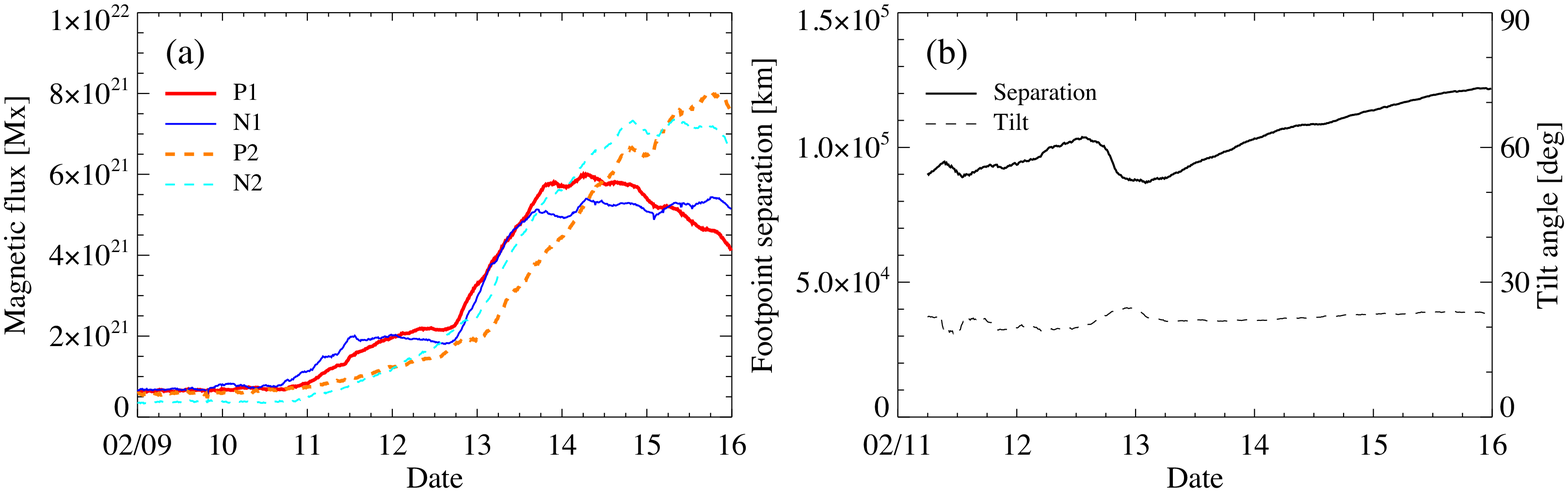}}
  \caption{(a) Total unsigned magnetic flux
    for each polarity.
    (b) Separation of the most distant polarities,
    P1 and N2,
    and the tilt angle between them.
  }
  \label{fig:ar}
\end{figure}

Figure \ref{fig:ar}a shows
the temporal evolution
of the total unsigned magnetic flux
of each polarity
in this AR.
One may see that
P1--N1 first increased its flux,
particularly from 10 February,
and P2--N2 increased
from 11 February.
Both bipoles continuously sourced their flux
from a series of minor emergences.
In this figure,
the total flux in each bipole
[P1--N1 and P2--N2]
is rather balanced
throughout the entire evolution.
However, in the final stage,
the fluxes of
both inner polarities,
N1 and P2,
are slightly larger
than their counterparts,
P1 and N2, respectively.
The total unsigned flux
of all four patches
reached
$2.5\times 10^{22}\ {\rm Mx}$.

In order to quantitatively describe
the geometrical evolution of this AR,
we measured the flux-weighted center
of each positive and negative polarity,
$(x_{\pm}, y_{\pm})$,
which is defined as
\begin{eqnarray}
  \left(
    x_{\pm}, y_{\pm}
  \right)
  = \left(
    \frac{\displaystyle\Sigma xB_{\pm}}{\displaystyle\Sigma B_{\pm}},
    \frac{\displaystyle\Sigma yB_{\pm}}{\displaystyle\Sigma B_{\pm}}
  \right),
\end{eqnarray}
where $B_{\pm}$ is
the field strength
of each pixel
and $+$ ($-$) is for
positive (negative) polarity.
Here,
only the pixels
with absolute field strengths
above 200 gauss
are considered.
Figure \ref{fig:ar}b shows
the separation of the outmost patches P1 and N2
and their tilt angle
{\it versus} time.
In our definition of the tilt,
the positive value indicates
that it agrees with Joy's tilt law
\citep{hal19};
the preceding (western) polarity
is on average closer to the Equator
than the following (eastern) part.
Therefore,
Figure \ref{fig:ar}b indicates
that the two patches gradually separated
from each other up to about $115\ {\rm Mm}$,
and that the tilt
was almost constant
with a final value of about
$+23^{\circ}$,
which matches Joy's law.
Here, the slight fall-off
of the separation
around 18:00 UT on 12 February
is due to another flux increment of P1,
which is also seen
in Figure \ref{fig:ar}a.

\subsubsection{Coronal Evolution}

\begin{figure}
  \centerline{\includegraphics[width=0.9\textwidth,clip=]{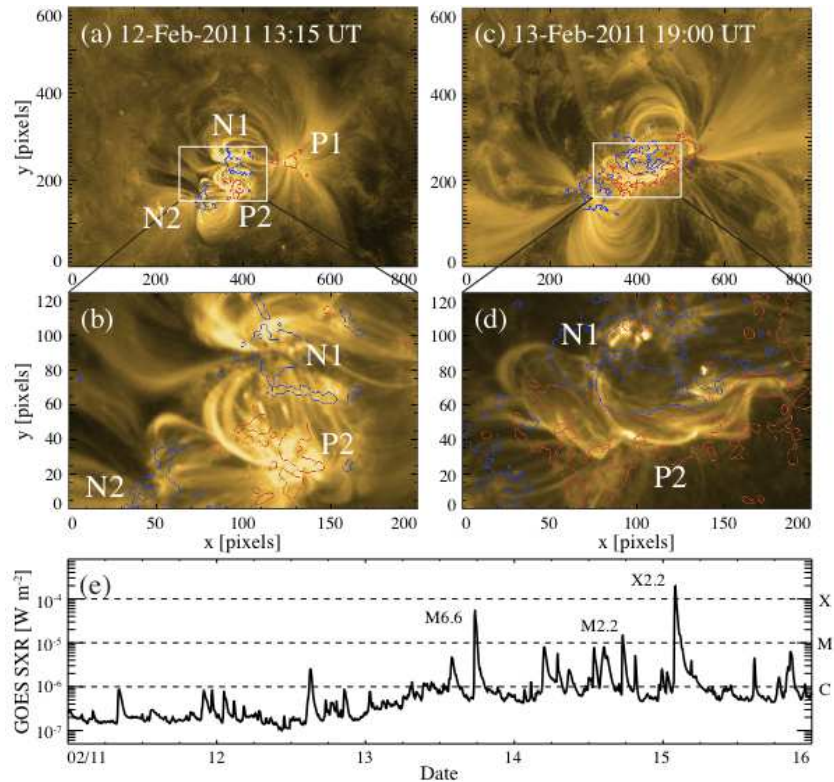}}
  \caption{Coronal evolution and flare events
    in AR 11158.
    (a) SDO/AIA 171 \AA\, image
    at 13:15 UT on 12 February 2011.
    One pixel corresponds to 0.6 arcsec.
    (b) Closeup of panel a.
    The field of view is shown in panel a.
    (c and d) Same as panels a and b
    but for 19:00 UT,
    13 February.
    (e) GOES-15 soft X-ray flux
    of $1.0$\,--\,$8.0$ \AA\, channel
    (five-minute cadence).
  }
  \label{fig:aia}
\end{figure}

In order to analyze the coronal structure,
we also used SDO/AIA data.
Figure \ref{fig:aia} shows
the evolution of the coronal
EUV structures
observed in 171 \AA.
As the two major bipoles,
P1--N1 and P2--N2,
appeared at the photosphere,
corresponding coronal loops
of the two bipoles
also rose into the corona.
One may notice that,
before N1 and P2
collided with each other
at 13:15 UT on 12 February
as in Figures \ref{fig:aia}a and b,
an arcade-like structure
connecting N1 and P2
was already created,
which may have been produced
by the magnetic reconnection
between the coronal loops
of the main bipoles
[P1--N1 and P2--N2].
Even when the photospheric patches
N1 and P2
are still separated,
the emerged magnetic fields
of the two bipoles
have already expanded
into the corona.
Between the expanding coronal loops,
a current sheet is formed
and eventually reconnection
starts to occur.

The creation of the arcade
suggests the existence
of the reconnected counterpart,
namely, overlying magnetic fields
that connect the most distant polarities:
P1 and N2.
However, we did not find
the clear P1--N2 loops
in the 171 \AA\, images.
This may be because
the intensity at the top of the long loop
is substantially weaker
than that at the footpoint.
If we assume that
the plasma within the loop
is in hydrostatic equilibrium
and its temperature is about $10^{6}\ {\rm K}$
({\it i.e.} the density scale height
is $\approx 50\ {\rm Mm}$),
the density at the top
of the loop
with a height of $50$\,--\,$100\ {\rm Mm}$
(1\,--\,2 times the scale height)
is several times smaller
than the footpoint.
Since the intensity is proportional
to the square of the density,
the intensity at the looptop
will be at least
one order of magnitude
weaker than the footpoint.
On 11 February,
we observed that
the coronal loop connecting P1--N1
suddenly expanded and became faint,
which may hint at
the reconnection
between P1--N1 and P2--N2
and the creation of P1--N2.

At the photosphere,
both inner polarities N1 and P2
continuously drifted
in a counterclockwise direction
with time.
Here, P2 moved to the West
and created a sheared PIL
between N1 and P2.
Figures \ref{fig:aia}c and d
show the corresponding coronal structure
at 19:00 UT on 13 February.
The arcade field N1--P2
is highly sheared,
following the footpoint motions
of the photospheric patches.
The series of flares in this AR
was observed
around this sheared PIL
after the PIL was formed
on 13 February.
Figure \ref{fig:aia}e
is the light curve
of GOES soft X-ray flux
($1.0$\,--\,$8.0$ \AA\, channel)
for five days.
The two strongest flares,
M6.6 and X2.2,
occurred at this PIL
on 13 and 15 February,
respectively
(see, {\it e.g.}, \opencite{liu12};
\opencite{sun12}).
Therefore,
we can conclude that
the production of intense flares
is strongly related to
the creation of a long, sheared PIL.
Note that some flares
were at different locations.
For example,
the M2.2 event
of 14 February occurred
much closer to the N2 polarity,
not at the central PIL.

\subsection{Summary and Interpretation
  of the Observations}

In this subsection,
we summarize the observation results.
NOAA AR 11158 appeared
on the southern hemisphere
in February 2011
as a quadrupolar region
composed of two major bipoles.
The eastern bipole [P1--N1]
first
appeared in
the HMI magnetogram
on 9 February
and then
the western pair [P2--N2]
emerged on the next day.
As their flux content increased,
the coronal fields also became visible
in the AIA images.
Before the two inner polarities
[N1 and P2] collided with each other
to form a $\delta$-configuration,
a coronal arcade
connecting these two polarities
was observed,
which was created
via magnetic reconnection
between expanded coronal loops
of both bipoles.
However,
magnetic fields arching over the entire AR
were not clear,
probably because the density
at the looptop
was much smaller
than the footpoint.
The continuous photospheric motion
of N1 and P2
strongly sheared the PIL
in the center
of this AR
and the coronal arcade
above that.
The existence of
the long, sheared PIL
between elongated polarities
and the coronal arcade
above the PIL
is of great importance
for producing
the train of strong flares
including the X2.2 and M6.6 events.
The total size of this AR
eventually reached
more than $150\ {\rm Mm}$
and the maximum unsigned flux
was $2.5\times 10^{22}\ {\rm Mx}$.
We found that
the fluxes of inner polarities
were larger than
those of outer polarities
in the final stage.
The centroids of the outer polarities
P1 and N2
separated up to $115\ {\rm Mm}$
with a tilt angle of $+23^{\circ}$,
which follows Joy's law.

\begin{figure}
  \centerline{\includegraphics[width=0.9\textwidth,clip=]{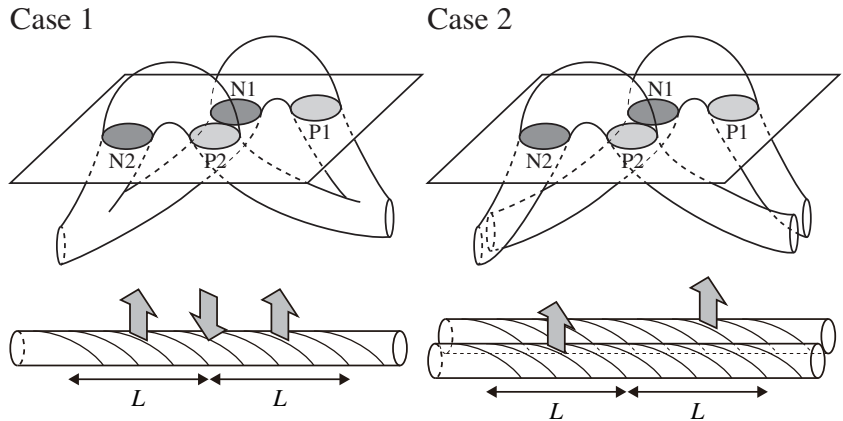}}
  \caption{(Top) Proposed concept
    of the flux system
    that forms AR 11158.
    Ellipses on the plane (photosphere)
    indicate the two major bipoles,
    P1--N1 and P2--N2.
    Lighter and darker shadows mean
    positive and negative polarities,
    respectively.
    Tubes above and below the photosphere
    show the expected flux tubes
    that compose this AR.
    Case 1 is the emergence
    of a single, split flux tube,
    while Case 2 is the emergence
    of two neighboring, independent tubes.
    (Bottom) Schematic illustration
    of the initial flux tubes
    in the numerical simulation
    for Cases 1 and 2.
    Arrows indicate
    the rising and sinking portions,
    while $L$ is the length
    for a field line
    to rotate once
    around the axis.
  }
  \label{fig:tube}
\end{figure}

Based on the above observation,
we propose models
for a global flux system
that creates such a sheared PIL
with a potential
to produce flares.
Figure \ref{fig:tube} shows
two possible models
for AR 11158.
Case 1 is the double flux emergence
from a single flux tube.
In this case,
an isolated tube is perturbed
in the convection zone
and split into two,
which rises to the surface
as two adjacent bipoles
[P1--N1 and P2--N2].
Case 2 is the emergence
of two different, but neighboring, tubes.
These tubes rise separately
to the surface at two locations.
In this case,
it may be possible that
the nearly simultaneous emergence
of the two bipoles
is triggered
by the same perturbation
({\it e.g.} convective upflow).
In the next section,
we carry out numerical simulations
to investigate which case is more realistic
as a possible scenario
for AR 11158.

\section{Numerical Simulations
  \label{sec:simulation}}

In order to explore
the global/subsurface structure
of AR 11158,
particularly to study
which of the magnetic systems
of Figure \ref{fig:tube}
could produce the highly sheared, long PIL
and the coronal arcade,
we conducted numerical simulations
of the flux tube emergence.
In the observations
in Section \ref{sec:observation},
these features were
found to be important
for the production
of intensive flares.

\subsection{Basic Setup}

The setup of the simulation
is basically the same as
that in \inlinecite{tor11}.
Here, we solve nonlinear,
time-dependent, fully compressible MHD equations.
First,
we set a 3D simulation box
$(-120,-120,-20)\le(x/H_{0},y/H_{0},z/H_{0})\le(120,120,150)$,
which is resolved by
$256\times 256\times 256$ grid points.
Here, $H_{0}$ is the normalizing unit
for length.
The grid spacings
in the middle of the domain
are $\Delta x/H_{0}=\Delta y/H_{0}=0.5$
for $(-40,-40)\le (x/H_{0},y/H_{0})\le(40,40)$
and $\Delta z/H_{0}=0.2$
for $-20\le z/H_{0}\le 20$.
The spacings outside this range
gradually increase.
Periodic boundaries are assumed
for both horizontal directions
and symmetric boundaries
for the vertical.
We define
the background stratification as
the adiabatically stratified convection zone
$[z/H_{0}<0]$,
the cool, isothermal photosphere/chromosphere
$[0\le z/H_{0}<10]$,
which we simply call the photosphere,
and the hot, isothermal corona
$[z/H_{0}\ge 20]$.
The photosphere and the corona
are smoothly connected
by a transition region.
Note that all of the physical values
are normalized by
$H_{0}=170\ {\rm km}$ for length,
$C_{\rm s0}=6.8\ {\rm km\ s}^{-1}$ for velocity,
$\tau_{0}=H_{0}/C_{\rm s0}=25\ {\rm seconds}$ for time,
and $B_{0}=250\ {\rm gauss}$ for magnetic-field strength.

We embed initial twisted flux tubes
in the convection zone.
In this study,
we test two cases:
Case 1 is a single, split flux tube
that emerges at two locations
(see Case 1 of Figure \ref{fig:tube}).
Here, we mimic
the splitting of the tube
by sinking the middle part.
Case 2 is the emergence of two different
but neighboring flux tubes
(see Case 2 of Figure \ref{fig:tube}).
The axial and azimuthal components
of a flux tube
are given as follows:
for a radial distance from the axis
$r=[(y-y_{\rm tube})^{2}+(z-z_{\rm tube})^{2}]^{1/2}$,
\begin{eqnarray}
  B_{x}(r)
  =B_{\rm tube}\exp{ \left( -\frac{r^{2}}{R_{\rm tube}^{2}} \right)}
\end{eqnarray}
and
\begin{eqnarray}
  B_{\phi}(r)=qrB_{x},
\end{eqnarray}
where $(y_{\rm tube},z_{\rm tube})$
denotes the tube center,
$R_{\rm tube}$ the typical radius,
$q$ the twist parameter
(uniform twist),
and $B_{\rm tube}$ the field strength at the axis.
We take $B_{\rm tube}/B_{0}=-15$,
$R_{\rm tube}/H_{0}=2.5$,
and $qH_{0}=0.2$
for all the tubes.
Here, the axial field
is directed to the negative $x$
in order to fit
to the expected toroidal field
in the southern hemisphere
in Solar Cycle 24
and, thus,
$B_{\rm tube}$ has a negative value.
Moreover,
these parameters mean that
the twist of the tubes is right-handed,
which is also
favorable to
the southern hemisphere
\citep{pev95}
and is stable
against the kink instability
\citep{lin96}.
The tubes are embedded at
$(y_{\rm tube}/H_{0},z_{\rm tube}/H_{0})=(0,-10)$
for Case 1
and $(\pm 5,-10)$ for Case 2.
In order to initiate their rise,
we perturb the tubes
by reducing densities
at the initial state.
The density reduction
has a function of
$\cos{(k_{x}x-\phi)}$,
where $k_{x}=2\pi/L$
and $L=2\pi/q=10\pi H_{0}$ is
the length for a field line
to make one helical turn
around the axis,
and $\phi$ is
the azimuthal angle
in the cross section
measured from $\hat{z}$.
Note that
this type of perturbation
is similar to that of the helical-kink instability
({\it e.g.} \opencite{fan98b}),
but the present tubes
are not kink-unstable,
at least in the initial condition.
In Case 1,
we limit the density reduction
to $-L\le x\le L$,
namely,
the tube is most buoyant at
$x=\pm L/2$.
In Case 2,
the reduction is limited
to $-L\le x\le 0$ for a tube
in the negative $y$
($y_{\rm tube}/H_{0}=-5$)
and to $0\le x\le L$
for a tube in the positive $y$
($y_{\rm tube}/H_{0}=5$).
The bottom panels
of Figure \ref{fig:tube}
illustrate
the initial perturbations
with arrows.

We also adopt anomalous resistivity
only over the range of
$-15\le x/H_{0}\le 15$
and $-10\le y/H_{0}\le 10$
to trigger the magnetic reconnection
in the center of the computational domain.
The resistivity will be switched on
when the current density
normalized by the plasma density
exceeds the threshold,
and it is a positive function
of the normalized current.
This promotes the reconnection
at the location
where the magnetic fields
are anti-parallel
and the density is lower
({\it e.g.} in the corona).
Outside the central domain
($|x|/H_{0}>15$, $|y|/H_{0}>10$),
the resistivity is always set to zero.

\subsection{Results}

\subsubsection{Photospheric Evolution}

\begin{figure}
  \centerline{\includegraphics[width=0.9\textwidth,clip=]{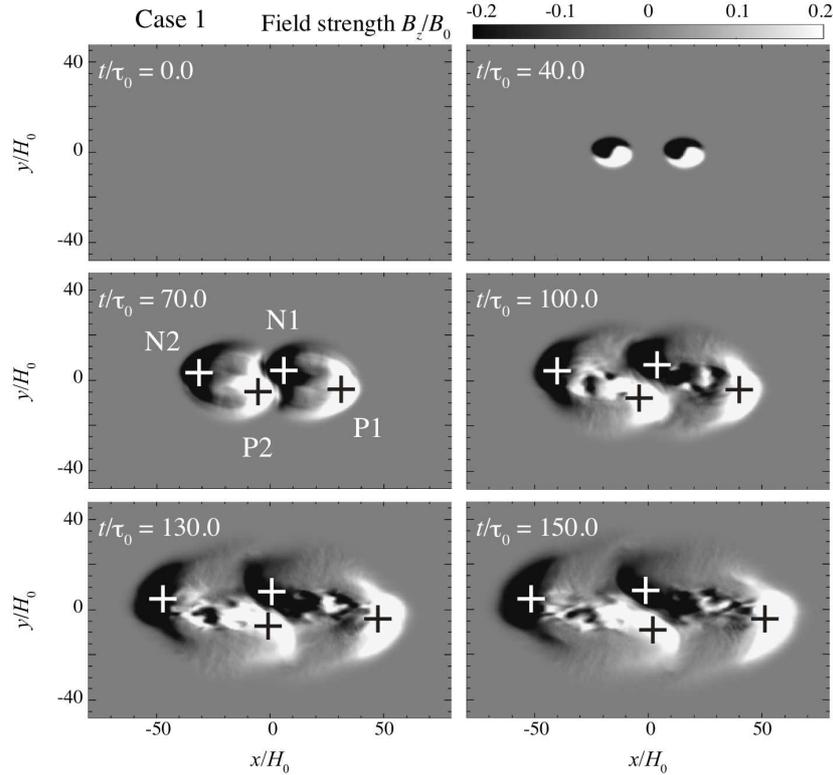}}
  \caption{Temporal evolution
    of the vertical fields $[B_{z}/B_{0}]$
    measured at the surface $[z/H_{0}=0]$
    (magnetogram)
    of the simulation Case 1.
    The flux-weighted centers
    of the four major polarities,
    P1, N1, P2, and N2,
    are indicated with crosses.
  }
  \label{fig:case1}
\end{figure}

\begin{figure}
  \centerline{\includegraphics[width=0.9\textwidth,clip=]{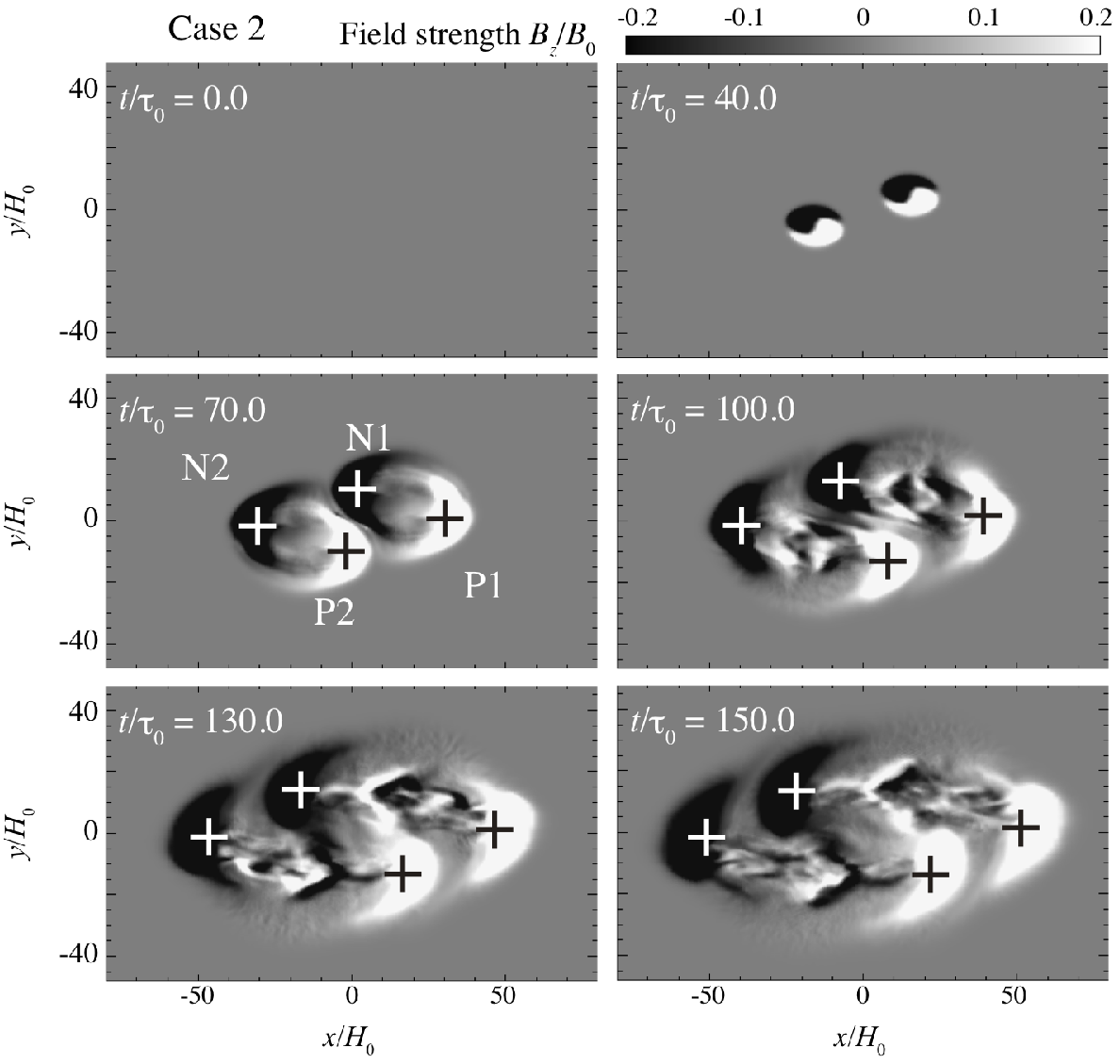}}
  \caption{Temporal evolution
    of the vertical fields $[B_{z}/B_{0}]$
    measured at the surface $[z/H_{0}=0]$
    (magnetogram)
    of the simulation Case 2.
    The flux-weighted centers
    of the four major polarities,
    P1, N1, P2, and N2,
    are indicated with crosses.
  }
  \label{fig:case2}
\end{figure}

Figures \ref{fig:case1} and \ref{fig:case2} show
the temporal evolution
of the vertical magnetic fields
$[B_{z}/B_{0}]$
measured at the photosphere
$[z/H_{0}=0]$
for Cases 1 and 2,
respectively.
In both cases,
the flux appeared
at the surface
at $t/\tau_{0}=32$
as two pairs
of positive and negative polarities,
which are named
P1--N1 and P2--N2 (from positive $x$).
As the pair of $\Omega$-shaped loops
rose into the atmosphere,
each polarity of the two bipoles
separated away from
its counterpart.

\begin{figure}
  \centerline{\includegraphics[width=0.65\textwidth,clip=]{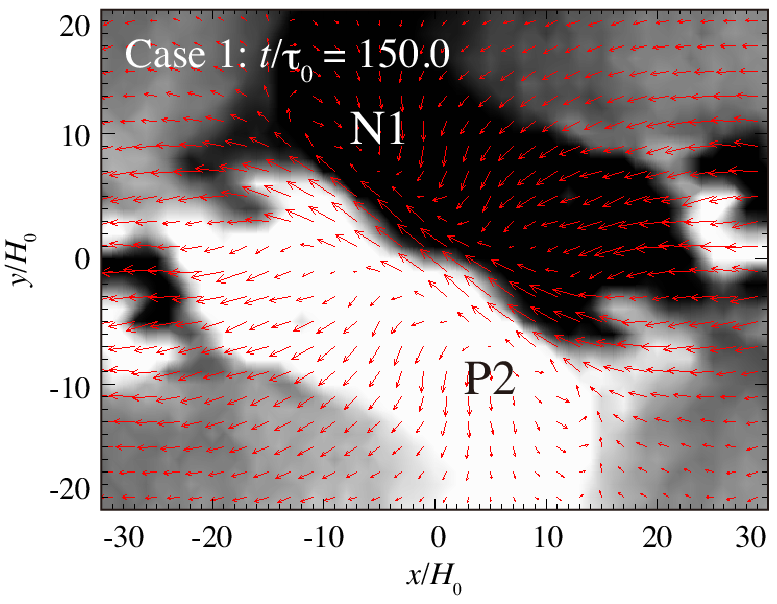}}
  \caption{Vertical and horizontal
    magnetic fields
    of Case 1
    at $z/H_{0}=0$ at $t/\tau_{0}=150$.
    The grayscale shows the vertical fields,
    which saturates at $B_{z}/B_{0}=\pm 0.1$,
    and the red arrows indicate
    the horizontal fields.
  }
  \label{fig:pil}
\end{figure}

In Case 1 in Figure \ref{fig:case1},
N1 and P2 approached each other
at around $t/\tau_{0}=70$
and created a PIL between them.
In this simulation,
the two emerging bipoles
belong to the same original flux tube,
and thus the two inner footpoints
[N1 and P2]
moved to the middle
of the original tube
[$x/H_{0}=0$: the sinking point].
Therefore,
a clear PIL
was built in the middle
of this quadrupolar region.
Figure \ref{fig:pil} shows
the vertical and horizontal fields
of Case 1
at the surface
at $t/\tau_{0}=150$.
One may see that
the horizontal field
on the PIL
is highly sheared
and is mostly parallel to the PIL,
due to the elongations
of N1 and P2.
Also, the clustering
of N1 and P2
is strongly reminiscent of
the formation of
$\delta$-sunspots.

Contrary to Case 1,
the two inner polarities
of Case 2,
as shown in Figure \ref{fig:case2},
did not show
any significant interaction.
At $t/\tau_{0}=70$,
N1 and P2
approached each other
at the middle of the region.
However,
they just passed by
at $t/\tau_{0}=100$
without forming a clear PIL
in between.
This is because the two emerging loops
originate from the two different flux tubes.
The photospheric footpoints
of these loops
just trace the axes
of their original tubes,
which align parallel
to each other
in the $x$-direction
and never cross
in the convection zone.
Therefore,
N1 and P2 continued
their migrations
even after they encountered
at the center of the region.
Finally, at $t/\tau_{0}=150$,
there were four isolated polarities
with no elongated PIL
in the magnetogram.

\begin{figure}
  \centerline{\includegraphics[width=1\textwidth,clip=]{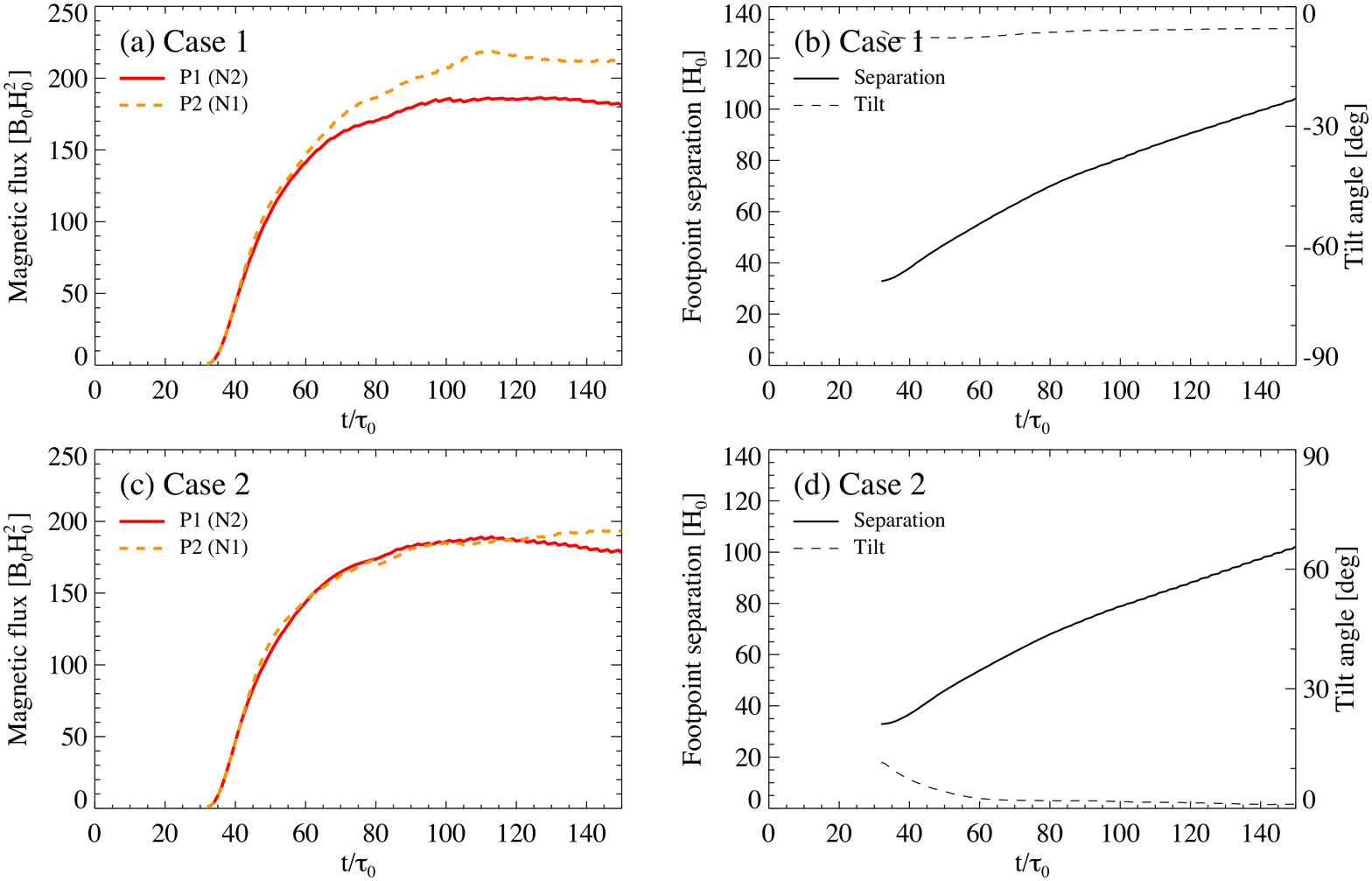}}
  \caption{(a) Total unsigned flux of P1 and P2
    in the simulation Case 1.
    The unsigned flux of N1 and N2
    is the same as that of P2 and P1,
    respectively,
    due to the symmetry of the simulation.
    (b) Separation between most distant polarities
    P1 and N2 of Case 1
    and their tilt angle.
    (c) The same as panel a
    but for Case 2.
    (d) The same as panel b
    but for Case 2.
  }
  \label{fig:sim}
\end{figure}

Figure \ref{fig:sim}a shows
total unsigned fluxes
of P1 (N2) and P2 (N1)
of Case 1.
Note that,
because of the symmetry of the simulation,
the unsigned flux of N1 and N2
is just the same as P2 and P1,
respectively.
From this figure,
we can see that
there is a flux imbalance
within each bipole
[P1--N1 and P2--N2].
Here,
the flux of the inner polarity
[P2 and N1]
is larger than that of its counterpart.
This discrepancy can be explained
by the following discussion.
Some low-lying
fields may undulate
around the photosphere
and penetrate the surface
at several sites.
Such serpentine field lines
were found in the previous simulations
({\it e.g.} \opencite{tor11}).
If field lines of inner polarities
are more undulated
as a result of the collision,
the total flux
of these polarities
may become larger,
which brings the flux imbalance
between inner and outer polarities.
Figure \ref{fig:sim}c shows that,
in Case 2,
the flux imbalance
within each bipole
is less obvious
than that in Case 1.
This may be because
minor interactions
of the inner polarities
in Case 2
causes less undulation
of the field lines.

The separations of outer polarities
P1 and N2
of Cases 1 and 2
and their tilt angles
are shown in
Figures \ref{fig:sim}b and d,
respectively.
In both cases,
the size of the AR continuously increased
with time.
At $t/\tau_{0}=150$,
the separations were
about $100H_{0}=17\ {\rm Mm}$.
The tilt in Case 1 was almost constant
and was about $-5^{\circ}$
from the $x$-axis,
while, in Case 2,
the tilt gradually decreased to $+1^{\circ}$.
Note that
our simulation models
ignored the tilt produced
by the Coriolis force
while the flux rises
through the convection zone
(\opencite{dsi93}).
The tilt here,
or, the deviation from the $x$-axis,
is simply caused by
the twist and the relative positions
of the original flux tubes.

\subsubsection{Coronal Evolution}

\begin{figure}
  \centerline{\includegraphics[width=0.8\textwidth,clip=]{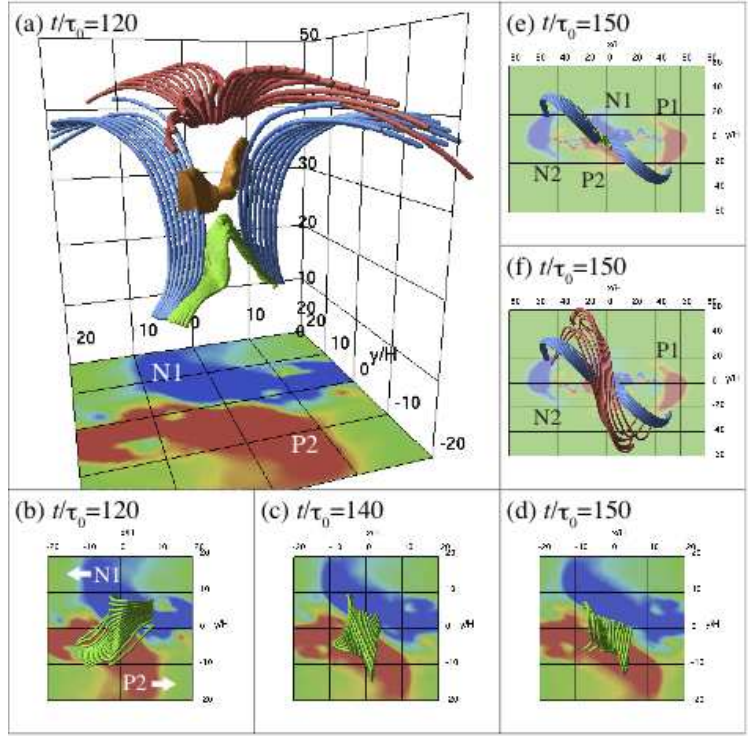}}
  \caption{Magnetic reconnection of the simulation Case 1.
    (a) Reconnection
    and the corresponding fields
    in the center of the simulation domain
    $(-20,-20)\le(x/H_{0},y/H_{0})\le(20,20)$
    at the time $t/\tau_{0}=120$.
    Field lines of the two emerging bipoles (blue) reconnect
    and form a low-lying arcade (green)
    and post-reconnection fields (red).
    The current sheet is shown
    by the isosurface (orange),
    which is
    plotted over $-5\le x/H_{0}\le 5$,
    $-5\le y/H_{0}\le 5$,
    and $z/H_{0}\ge 26$.
    Reconnection occurs at $z/H_{0}\approx 26$.
    Photospheric vertical field $B_{z}/B_{0}$
    is plotted at the bottom,
    whose color scale saturates
    at $0.2$ (red) and $-0.2$ (blue).
    (b\,--\,d) Evolution of the arcade field (green)
    from $t/\tau_{0}=120$ to 150.
    Photospheric motions of N1 and P2
    are indicated with arrows in panel (b).
    (e and f) Top view of the field lines
    without and with the post-reconnection fields (red)
    at $t/\tau_{0}=150$.
  }
  \label{fig:bvct}
\end{figure}

In this section,
we show the magnetic reconnection
between the coronal fields
and its evolution.
Figure \ref{fig:bvct}a is the closeup
of the reconnection site
of Case 1 for
$(-20,-20)\le(x/H_{0},y/H_{0})\le(20,20)$
at time $t/\tau_{0}=120$.
Here, the field lines of two emerging bipoles
P1--N1 and P2--N2
came close to each other,
and a current sheet
was formed in between.
Magnetic reconnection was triggered
in this current sheet
because of the anomalous resistivity
applied to the center
of the domain.
As a result,
a coronal arcade
was produced
above the photospheric PIL.
Figures \ref{fig:bvct}b\,--\,d
show the temporal evolution
of the arcade field
from $t/\tau_{0}=120$ to $150$.
Following the counterclockwise motion
of the photospheric patches N1 and P2,
the coronal arcade
above the PIL
was also sheared and rotated
in the counterclockwise direction.
At higher altitudes,
another flux system
was created
by magnetic reconnection
(post-reconnection fields).
This flux was ejected
from the reconnection site
with an upflow
of the order
of the local Alfv\'{e}n speed
($2$\,--\,$6C_{s0}$).
Figures \ref{fig:bvct}e and f
show the connectivity
of the large-scale fields
in the final state
at $t/\tau_{0}=150$.
From these figures,
one can see that
the emerging loops
connect the two photospheric polarities
of each bipole,
while the post-reconnection fields
link the most distant elements.

The coronal structure 
of simulation Case 2
is similar to
that of Case 1:
Both, the coronal arcade [N1--P2]
and the post-reconnection fields [P1--N2]
are formed
by magnetic reconnection
of the two emerging loops
[P1--N1 and P2--N2].
However,
since the photospheric patches
N1 and P2
in Case 2
move away from each other
very quickly
and without profound interaction
(see Figure \ref{fig:case2}),
the apparent rotational motion
of the coronal arcade occurs
on a much shorter time scale
than that in Case 1.

\subsection{Summary of the Simulations}

In this section,
we carried out two types
of flux emergence simulation.
The initial conditions
were based on
the proposed models
of AR 11158.
The first case was
the emergence of a single flux tube
that emerges in two portions,
while, in the second case,
we involved the emergence
of two parallel tubes.
We observed
the double bipolar emergence
in both cases.
However,
the creation of a highly sheared, elongated PIL
between N1 and P2
in the center of the region
was found only in Case 1.
Here, in Case 1,
the common subsurface root
between the two emerging bipoles
led to the assembly
of the inner polarities,
producing a packed cluster
with a magnetic configuration
resembling $\delta$-sunspots.
The significant interaction
between the two inner polarities
results in the flux imbalance
within each bipole.
On the other hand,
in Case 2,
the interaction
between the inner footpoints
of the two horizontal tubes
was not clear.
These polarities
just passed by each other
without forming a long sheared PIL,
which may be reflected
in the more balanced flux
within each bipole.

As time progressed,
the two expanded coronal fields
came close to each other
and, due to the resistivity,
magnetic reconnection
was provoked
at the center of the region.
Reconnection occurred
in both simulations.
The emerging loops
of both bipoles
built a current sheet in between
and reconnection proceeded
to create the low-lying arcade
connecting the inner polarities
and post-reconnection fields
joining the outmost polarities.
We also observed
the rotation of coronal arcade [N1--P2]
in both simulations,
along with the counterclockwise shearing
of the photospheric patches.

\section{Comparison between Observations and Simulations
  \label{sec:comparison}}

In this article,
we first analyzed
the observations
of NOAA AR 11158
obtained by SDO/HMI and AIA
to study the evolution
of this AR.
The key features
that we found in AR 11158
were i) the long sheared PIL
between the elongated magnetic elements
of opposite polarity,
N1 and P2,
that formed $\delta$-sunspots
in the middle of the region
and ii) the sheared coronal arcade
above this PIL
created by magnetic reconnection
between $\Omega$-loops
of two major bipoles.
The strong flares were repeatedly
produced at this PIL.
Based on the observations,
we proposed two scenarios
for the creation of AR 11158:
Case 1 is the emergence
of a single split tube
that emerges at two portions,
while Case 2 is the emergence
of two different neighboring tubes.

\begin{figure}
  \centerline{\includegraphics[width=1.0\textwidth,clip=]{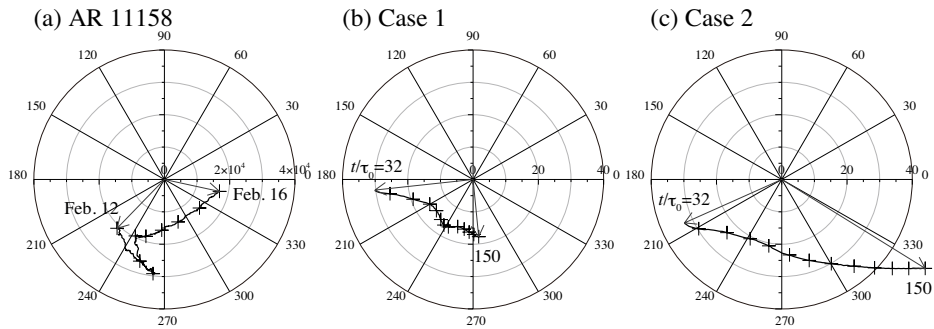}}
  \caption{Relative motion
    of the photospheric polarities
    N1 and P2
    for (a) AR 11158,
    (b) simulation Case 1,
    and (c) Case 2.
    The center of each diagram
    indicates the position of N1
    and the horizontal axis is parallel
    to the $x$-axis.
    (a) The plot is from 00:00 UT
    of 12 February 2011,
    to 00:00 UT,
    16 February.
    The crosses are plotted
    every 12 hours
    and the unit for length
    is $1\ {\rm km}$.
    (b and c) The plot is
    from $t/\tau_{0}=32$ to 150.
    The crosses are plotted
    every $10\tau_{0}$
    and the unit for length
    is $H_{0}=170\ {\rm km}$.
  }
  \label{fig:n1p2}
\end{figure}

Then, we conducted
flux-emergence simulations
of this two cases
to model AR 11158.
As a result,
Case 1 clearly revealed
a long, sheared PIL
and adjacent elongated polarities
N1 and P2,
which is well in line
with the observational results
of AR 11158.
However, in Case 2,
these polarities
just passed by each other
and never built a clear PIL,
which deviates
from the observation.
This discrepancy can easily be seen
in the diagrams
of Figure \ref{fig:n1p2}.
Here, we plot the temporal evolution of 
the direction and distance
from the centroid of N1
to that of P2,
{\it i.e.} the vector
from N1 to P2.
The center of the plot
corresponds to the position of N1.
This diagram was previously
introduced by \inlinecite{poi13}:
see Section \ref{sec:discussion}.
In AR 11158,
P2 continuously drifts
along the southern edge of N1
from East to West
in the counterclockwise direction.
Here, the distance
between N1 and P2 continuously decreases
as the shear becomes stronger.
Note that this shrinking of the vector
is not caused
by the change of viewing angle
associated with the solar rotation,
since we applied Postel's projection
to the tracked HMI magnetogram.
Between the two simulation results,
only Case 1 shows
a similar rotating and shrinking behavior.
During its anticlockwise rotation,
P2 gradually gets closer to N1,
creating a steep PIL.
The difference between
Figures \ref{fig:n1p2}a and b
is mainly due to the assumptions
in the simulation
(small numerical domain,
neglected Coriolis force, {\it etc}.).
For example,
the separation and the tilt
of P1--N2
are $115\ {\rm Mm}$ and $+23^{\circ}$
in AR 11158,
while those values are
$100H_{0}=17\ {\rm Mm}$
and $-5^{\circ}$
in simulation Case 1.
Thus, magnifying
Figure \ref{fig:n1p2}b 6.8 times
and rotating it $+28^{\circ}$
to match Figure \ref{fig:n1p2}a,
one may find clearer consistency
between the two figures.
However,
in Case 2 in Figure \ref{fig:n1p2}c,
P2 approaches N1
only in the initial phase.
The two polarities
simply fly by
and the distance becomes larger
at a constant pace,
which is opposite
to the observed behavior
of AR 11158.

Regarding the flux evolution
in the photosphere,
an imbalance
between inner and outer polarities
was observed
both in AR 11158
and in the simulations.
The fluxes of inner polarities
were larger than
those of outer polarities
in the final stage.
At least for the simulation,
the excess of the inner flux
can be explained
by the fluctuation
of the field lines
around the surface,
possibly due to the collision
between the two polarities.

In both simulation cases,
we found magnetic reconnection
and the creation of the arcade field
lying over the central PIL.
The arcade structure becomes sheared
following the motion
of the photospheric patches.
Although we did not find
clear post-reconnection fields
in the AIA images,
which are expected to connect
the most distant polarities
P1 and N2, 
the numerical results suggest
the existence of such magnetic fields.

From the above comparison,
we conclude here that
the flare-productive quadrupolar AR 11158
is,
between the two cases,
more likely to be composed
of a single flux tube
than two independent tubes.
The single tube
is severely perturbed
and separated into two
in the convection zone.
Then,
these two portions
appear at the photosphere
as two major bipoles.
Since the bipoles
share a common root
beneath the surface,
they recover
their original configuration
while the entire flux system
is expanding
into the corona,
where the gas pressure
is less dominant.
Thus the inner two polarities
show a strong
% confliction
gathering
and build a highly sheared PIL.
The coronal arcade
is also sheared
in that process,
which results in
the series of intense flares
including X- and M-class events.
If this AR
% were made up of
were composed of
two entirely separate magnetic bundles,
the inner polarities
cannot create
such a sheared PIL,
although the coronal arcade
above the PIL
might be built
through the magnetic reconnection
between coronal loops.

\section{Discussion
  \label{sec:discussion}}

As we discussed
in the previous section,
the SDO observations
and the flux-emergence simulations
suggest that,
between the two possible scenarios
of Figure \ref{fig:tube},
the flare-productive quadrupolar AR 11158
is more likely to be made
from a single subsurface flux tube.
That is,
the two emerging bipoles
at the photosphere
share a single large-scale structure
in the deep convection zone.
In the process of flux emergence,
they recover their original shape
by releasing the magnetic free energy
stored within the sheared coronal arcade
in the form of solar flares.

\inlinecite{poi13} observed
the photospheric and coronal evolution
of AR 10314.
This flare-productive quadrupolar AR
is very similar to AR 11158
in many aspects.
Their interpretation was that
AR 10314 is also created
by a single flux system.
Although they thought of
a rising $\Omega$-loop
whose top is curled downward
(see Figure 7 of \opencite{poi13}),
the basic concept
that the single flux system
emerges at two locations
is in line
with our study.
They argued that
the packed inner polarities
in the central AR
would not be created
by independent flux tubes.
Our simulation Case 2
has revealed that
their prediction is indeed true.
\inlinecite{chi13} analyzed AR 11158
and argued that
this AR originates from
a single flux tube
deformed in the convection zone,
based on the vertical stacking
of the sequential HMI magnetograms
(see Figure 2 of \opencite{chi13}).
Their concept is basically
consistent with our conclusion,
{\it i.e.} Case 1.

In the numerical simulations,
the initial depth
of the flux tubes
was assumed to be
$z_{\rm tube}=-10H_{0}=-1.7\ {\rm Mm}$.
However, it is likely
that the magnetic fields
that build ARs
are transported
from much deeper
in the convection zone.
Also, in the simulations,
the final separation
of two outer polarities
reached $100H_{0}=17\ {\rm Mm}$,
which is one order of magnitude
smaller than the $115\ {\rm Mm}$
observed in AR 11158.
\citeauthor{tor12} (\citeyear{tor12,tor13a})
may give us a suggestion
for the situation
when we expand the simulation box.
They found that,
even when the initial flux tube
is located at $-20\ {\rm Mm}$,
{\it i.e.}
ten times deeper
than the present simulations,
the flux tube indeed
reaches the surface
to form an AR
as long as it has
enough field strength,
total flux, and twist.
Therefore,
we can expect
that the simulation results
may hold,
at least to the depth of $20\ {\rm Mm}$.

In the simulations
of this study,
we did not find flare eruptions.
The main reason is that
a flux rope is not created
after the coronal arcade is built
through the magnetic reconnection
between the emerging loops
(see Figure \ref{fig:bvct}).
According to \inlinecite{kus12},
in order to create a flux rope
from a sheared arcade,
a small-scale flare-triggering magnetic field
is required,
which initiates
the reconnection between
the coronal arcades
and successfully produces
flaring activity.
In their simulation,
the small-scale field
that triggers the flare
is the result of an emerging flux
injected from the bottom boundary,
of which the origin has not been addressed
in their work.
In our calculation,
however,
such small-scale features
are not resolved.
Also,
according to the observation of
the M6.6-class event in AR 11158
by \inlinecite{tor13c},
the formation of such a small trigger
seems to be coupled with
local convection.
\inlinecite{bam13} analyzed
the flare triggers
in many more flaring events.
Since our calculation
does not include
thermal convection,
it may be difficult
to create a triggering field.
Therefore,
in our simulation,
the flux rope is not
formed
through the reconnection
between the coronal arcades,
resulting in no flare eruptions.

Here, we comment
on the previous simulations
of the flux-tube emergence
similar to our cases.
\inlinecite{mag07} simulated
emergence of a single flux tube
that rises at three portions.
The concept looks
similar to our Case 1.
However,
the aim of his simulation
was to investigate
the formation of a solar prominence.
Multiple emergence from a single flux system
and reconnection among these emerging loops
can be found
in a series of resistive-emergence simulations
\citep{iso07,arc09}.
The concept
of the present Case 2,
{\it i.e.}
the interaction between two neighboring flux tubes
in the atmosphere,
is seen in \inlinecite{arc07},
although their simulation was in 2D.
\inlinecite{gon09}
simulated the 3D emergence
of a toroidal flux tube,
not a horizontal tube,
and its reconnection
with a preexisting flux.
They found that a slingshot reconnection
produces an upward jet eruption
\citep{arc10}.
In our simulations,
we also observed
a similar upflow
from the reconnection site.
\inlinecite{arc13} simulated
the emergence of
a weakly twisted horizontal tube
and found an eruptive interaction
between the two expanding lobes,
which seems to be
similar to our Case 1.
The main difference
from our work is that
they calculated the emergence
initially at a single location.
After the flux
appears at the surface,
however,
it rises further
at two neighboring places,
resulting in the dynamical interaction
between the two lobes.
Contrary to this,
in our simulation Case 1
the initial emergence occurs at two places
and thus two expanding lobes are built,
which reconnect to form the coronal arcade.
Another interesting result
interaction
between the loops
in Case 1
is that
$\delta$-sunspots are created
from a kink-stable flux tube
(see, {\it e.g.}, \opencite{mat98}; \opencite{fan98b}).
Interaction of the flux tubes
within the convection zone
has also been studied in detail
({\it e.g.} \opencite{fan98a};
\opencite{lin01}; \opencite{mur07}).
In our Case 2,
there was no major interaction
between the two tubes
in the convection zone,
since they were initially
aligned in a parallel manner,
and they did not expand much
before they reached the photosphere.
However,
if we start the simulation
from the deeper convection zone,
or, if we consider a more complex alignment
of the initial tubes,
we need to take into account
the tubes' interaction
beneath the surface.

\section{Conclusion
  \label{sec:conclusion}}

In this study,
we compared the SDO/HMI and AIA observations
of NOAA AR 11158
with the numerical simulations
of the flux tube emergence,
aiming to study
the formation
of this flare-productive AR.
AR 11158,
basically composed of two emerging bipoles,
is one of the most flare-productive ARs
in Solar Cycle 24.
SDO's continuous observation
of the full solar disk
at multiple wavelengths
allows us to investigate
the photospheric and coronal
evolution of this AR
from the first minutes
to the moments of strong flares.
The key features
of this AR
that we found are
i) the long sheared PIL
between the elongated magnetic elements
of opposite polarity
that produced a $\delta$-configuration
in the central region
and ii) the sheared coronal arcade
above the PIL
created by magnetic reconnection
between $\Omega$-shaped loops
of two major bipoles.
Based on the SDO observations,
we inferred
two possible scenarios
for the formation
of AR 11158:
emergence of a single split tube
{\it versus} two different tubes.
According to the numerical simulations,
AR 11158 is more likely
to be made by a single flux tube
than two tubes.
The single tube is severely deformed
and split in two
while in the convection zone.
The two major bipoles
at the photosphere,
which share the common subsurface structure,
recover their original shape
during flux emergence
by releasing magnetic free energy
within the sheared coronal arcade
in the form of solar flares
including X- and M-class events.

Our study suggests that
a solar flare in an AR
naturally reflects
the large-scale magnetic structure
beneath the surface.
From this point of view,
flares can be thought
as a process
to relax magnetic shear
produced by a helical and/or deformed emerging flux
from the solar interior.
However,
we cannot observe
the subsurface structure
of flare-productive ARs
from direct optical observations.
Helioseismic detection
of the emerging subsurface flux
({\it e.g.} \opencite{ilo11};
\opencite{tor13b})
may improve our understanding
of the nature of such ARs.
Regarding the numerical simulations,
neither case
reproduced
flare reconnection,
since the evolution
of the flare trigger,
which may be coupled
with thermal convection,
was beyond the scope
of the simulation code.
Also,
the origin of
the perturbation that splits
the emerging flux in two,
which was mimicked
by the sinking of the tube,
remained unclear.
These topics will be addressed
in future investigations.

%%%%%%%%%%%%%%%%%%%%%%%%%%%%%%%%%%%%%%%%%%%%%%%%%%%%%%%%%%%%%%%%%%%%%%%%%%%
%% Appendix
%
% \appendix   

%%%%%%%%%%%%%%%%%%%%%%%%%%%%%%%%%%%%%%%%%%%%%%%%%%%%%%%%%%%%%%%%%%%%%%%%%%%
%% Acknowledgements
%
\begin{acks}[Acknowledgments]
The authors thank
K. Hayashi (Stanford University)
for providing HMI data
and P. D\'{e}moulin
(Paris Observatory)
for fruitful discussions.
The authors are grateful
to the anonymous referee
for improving the manuscript
and the SDO team
for distributing HMI and AIA data.
Numerical computations were carried out
on Cray XC30
at the Center for Computational Astrophysics,
National Astronomical Observatory of Japan.
ST is supported by
Grant-in-Aid for JSPS Fellows.
This work was supported
by a Grants-in-Aid for Scientific Research (B)
``Understanding and Prediction of Triggering Solar Flares''
(23340045, Head Investigator: K. Kusano)
from the Ministry of Education, Science, Sports,
Technology, and Culture of Japan.
\end{acks}

%%% %%%%%%%%%%%%%%%%%%%%%%%%%%%%%%%%%%%%%%%%%%%%%%%%%%%%%%%%%%%
%% Bibliography
%
% Using BibTeX
%
\bibliographystyle{spr-mp-sola}
\tracingmacros=2
\bibliography{toriumi2014}  
\IfFileExists{\jobname.bbl}{} {\typeout{}
\typeout{****************************************************}
\typeout{****************************************************}
\typeout{** Please run "bibtex \jobname" to obtain} \typeout{**
the bibliography and then re-run LaTeX} \typeout{** twice to fix
the references !}
\typeout{****************************************************}
\typeout{****************************************************}
\typeout{}}
%
% Without BibTeX 
% \begin{thebibliography}{}
% \bibitem[\protect\citeauthoryear{Author}{Year}]{key}
%   <bibliographical entry>
%
% \bibitem[\protect\citeauthoryear{}{}]{}
%   
%  
% \end{thebibliography}

\end{article} 
\end{document}